\documentclass{moriond}

\def\Journal#1#2#3#4{{#1} {\bf #2}, #3 (#4)}

\def\PLB{{\em Phys. Lett.}  B}

\def\IJMP{{\em Int. J. Mod. Phys.} A}
\def\EPJ{{\em Eur. Phys. J.} C}
\def\JHEP{{\em JHEP}}

\newcommand{\Photo}{\includegraphics[height=35mm]{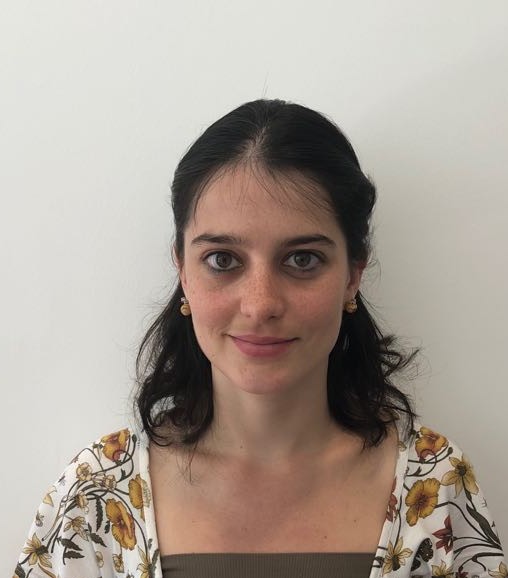}}

\begin{document}
\vspace*{4cm}
\title{MEASUREMENT OF JETS AND PHOTONS WITH THE ALICE EXPERIMENT AT THE LHC}
\author{L. LEARDINI }
\address{Physikalisches Institut, Heidelberg University, Im Neuenheimer Feld 226,\\
69120 Heidelberg, Germany}

\maketitle\abstracts{
Jets and direct photons are used to probe the formation and the properties of 
the Quark-Gluon Plasma (QGP) by the ALICE Collaboration. The results shown here will present the measurements 
obtained from Pb--Pb and p--Pb collisions. It is observed that in Pb--Pb collisions 
jets are collimated and hadron yields ($\pi^{0}$ and $\eta$) at high transverse momentum 
undergo strong suppression due to the interaction of the patrons with the QGP. On the other hand, 
photons do not show any suppression in this region, and, in addition, an excess attributed 
to a thermal emission is observed at low transverse momentum. No hints of medium modifications 
are observed for the p--Pb results, for which a jet quenching limit of $-$0.4~GeV/$c$ is calculated.}

\section{Introduction}
The main focus of the ALICE experiment is the study of the ultra-relativistic heavy-ion collisions 
and of the subsequent hot and dense phase of the matter, the Quark-Gluon Plasma (QGP). The 
understanding and characterisation of the properties of the created medium is accomplished 
using, among others, hard probes: the prompt component of the direct photons and jets from 
parton fragmentation. Prompt photons and jets have a common origin in the hard scatterings 
in the initial phase of the system evolution after the initial nucleus-nucleus collision. 
The partons created in the initial hard collisions scatter and lose energy while moving in 
the QCD medium. The modification of the hadron yields originating from these partons as a 
function of momentum gives information on the medium. This quantity is studied for several 
particle species, where the comparison to pp and p--Pb results contributes to the 
characterisation of the parton energy loss and QGP evolution. While for jet measurements the 
attention is focused on the modifications induced by the medium, a key feature of the direct 
photon (\emph{i.e.}~not coming from hadron decays) measurement is that its information is 
unaffected by the passage through the QCD matter. However, the medium itself will be a source 
of thermal emission and will lead to additional contribution of direct photons at low transverse 
momentum. Thus, the early stages are accessible in the form of the direct photon emission 
temperature and collective flow measurements. Considering that most of the photons are created 
in hadron decays, the direct photon measurement requires a good knowledge of the decay photon 
contribution to the inclusive photon yields.

\section{Jet measurements}
\begin{figure}[!h]
\begin{minipage}{0.55\textwidth}
\vspace*{-1.9cm} 
\includegraphics[width=0.8\textwidth]{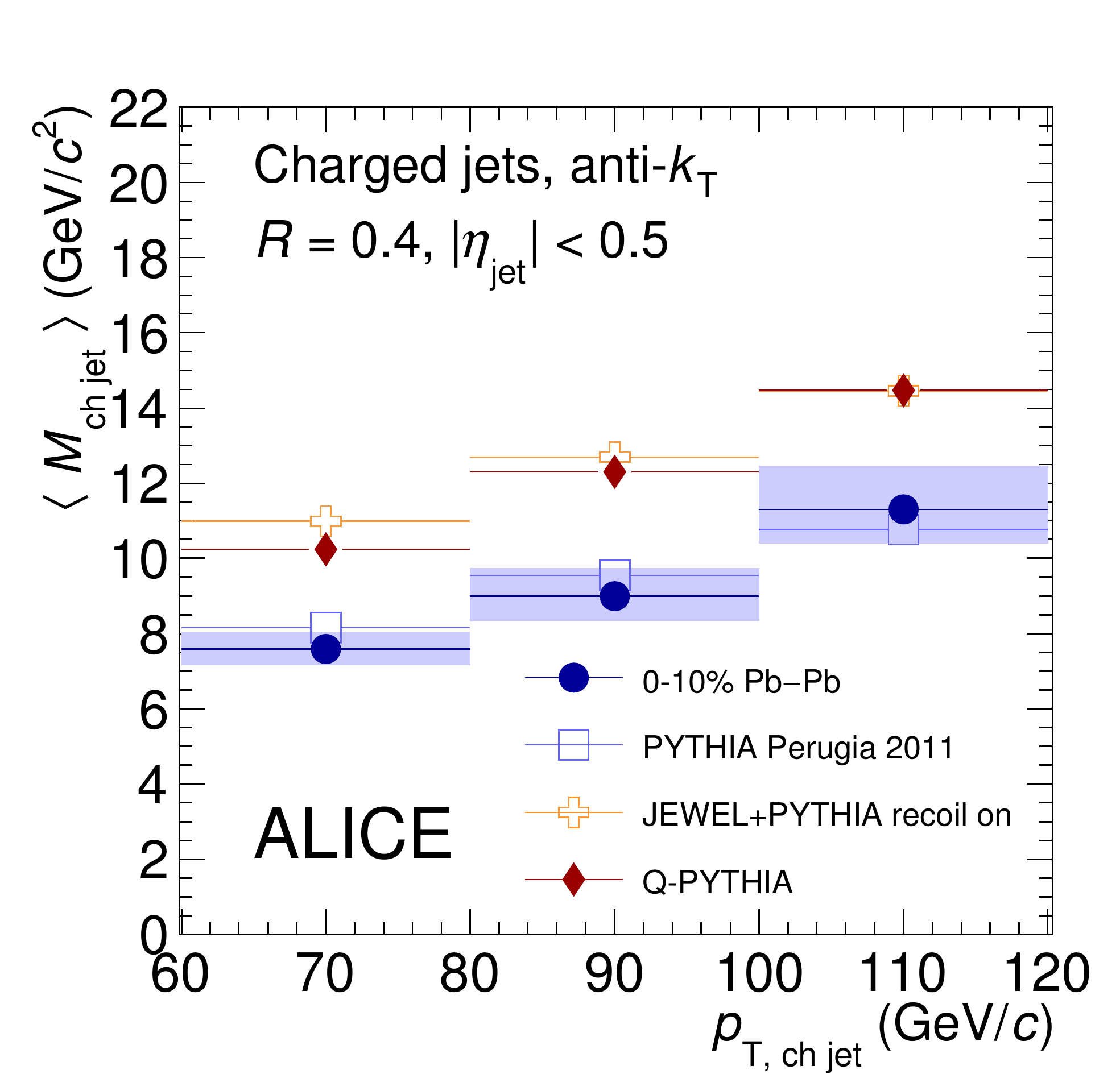}
\end{minipage}
\begin{minipage}{0.55\textwidth}
\includegraphics[width=0.8\textwidth]{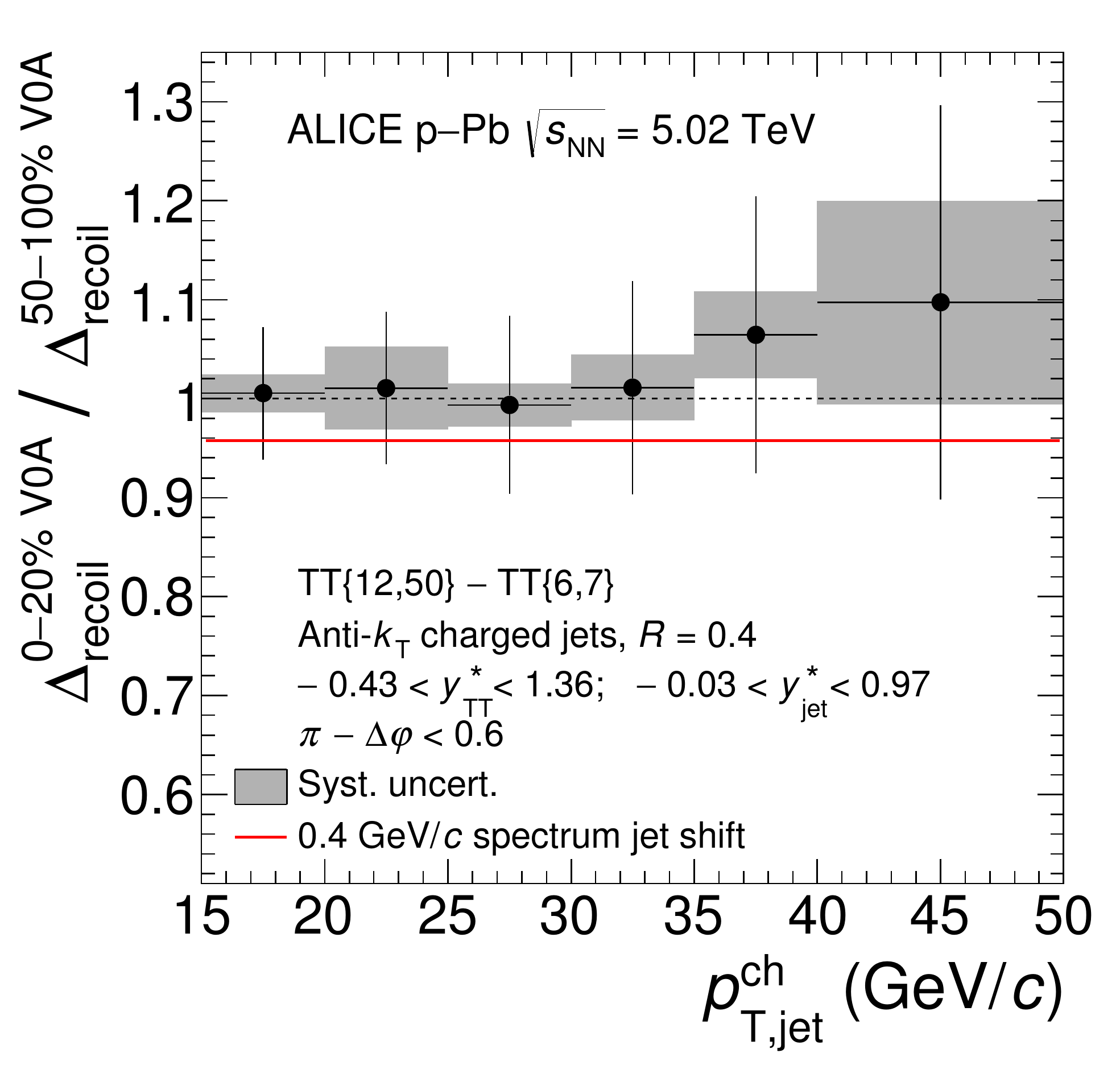}
\vspace*{1.2cm} 
\end{minipage}
\vspace*{-1.5cm}
\caption{Left: Fully-corrected mean jet mass compared to Pythia and to event generators 
with jet quenching (JEWEL and Q-PYTHIA) for jets with $R$~=~0.4 in the 10\% most central 
Pb--Pb collisions at $\sqrt{s_{\mbox{\tiny NN}}}$~=~2.76~TeV~$^{3}$. Right: Ratio of 
$\Delta_{\mbox{\tiny recoil}}$ distributions measured with $R$~=~0.4 for 0--20\%/50--100\% 
in p--Pb collisions at $\sqrt{s_{\mbox{\tiny NN}}}$~=~5.02~TeV~$^{5}$. The grey boxes show the 
systematic uncertainty of the ratio, including the correlated uncertainty of numerator 
and denominator. The red line indicates the jet quenching limit as a 
$p_{\mbox{\tiny T}}$-shift of $-$0.4~GeV/$c$.}
\label{fig:JetMeas}
\end{figure}

In ALICE, jets can be measured via charged-particle tracking, using the central barrel detectors, and 
via clusters in the EMCal, one of the electromagnetic calorimeters~\cite{Abelev:2014ffa}.
Jets are reconstructed using FastJet: the anti-$k_{\mbox{\tiny T}}$ algorithm is used for 
signal jets while the $k_{\mbox{\tiny T}}$ algorithm is used for the average background estimation. The 
typical resolution parameters are $R$~=~0.2 and 0.4. The ALICE Collaboration already measured jet quenching 
in Pb--Pb collisions~\cite{Adam:2015ewa} but to better characterise the medium properties, 
detailed studies on jet shapes are carried out. \\
In Fig.~\ref{fig:JetMeas} (left), the fully-corrected mean jet mass as a function of the 
charged jet $p_{\mbox{\tiny T}}$ measured in 0--10\% most central Pb--Pb collisions at 
$\sqrt{s_{\mbox{\tiny NN}}}$~=~2.76~TeV centre-of-mass energy~\cite{Acharya:2017goa} is shown. 
The jet mass is directly related to the leading parton virtuality and it can be used as a 
measure of the jet broadening: soft splittings within the jet cone will increase the jet 
mass while large angle radiation outside the jet cone will reduce the jet mass. The measurement 
points to a collimated jet 
structure, with the mean value of the jet mass increasing for increasing jet transverse momentum, 
as expected from NLO pQCD calculations. Moreover, the comparison to models with and without jet quenching 
indicates that competing effects, energy loss in the medium and medium response, enter this 
measurement and that further efforts are needed in order to isolate them. \\
The study of p--Pb collisions, that started out as a way to disentangle initial and final state and 
cold nuclear matter effects, became of relevance by itself to verify whether QGP formation 
takes place also in small systems. As mentioned above, the jet quenching has already been 
measured in central Pb--Pb in ALICE, setting the lower limit at $-$8~GeV/$c$ of jet spectrum 
shift~\cite{Adam:2015doa}. The same measurement has been carried out in p--Pb collisions using 
semi-inclusive hadron-jet coincidence~\cite{Acharya:2017okq}. This method allows for an unbiased 
measurement, independent of the collision geometry. The observable considered is the ratio of the 
recoil distributions ($\Delta_{\mbox{\tiny recoil}}$) in high (0--20\%) and low (50--100\%) multiplicity collisions, 
shown in Fig.~\ref{fig:JetMeas}, (right). In the presence of jet quenching, the ratio should be 
below unity. As no suppression is observed, the limit for jet quenching in small systems is 
set as a $-$0.4~GeV/$c$ shift in the jet spectrum, indicated by the red line in the figure.

\section{Neutral mesons and photon measurements}
\begin{figure}[!h]
\includegraphics[width=0.333\textwidth]{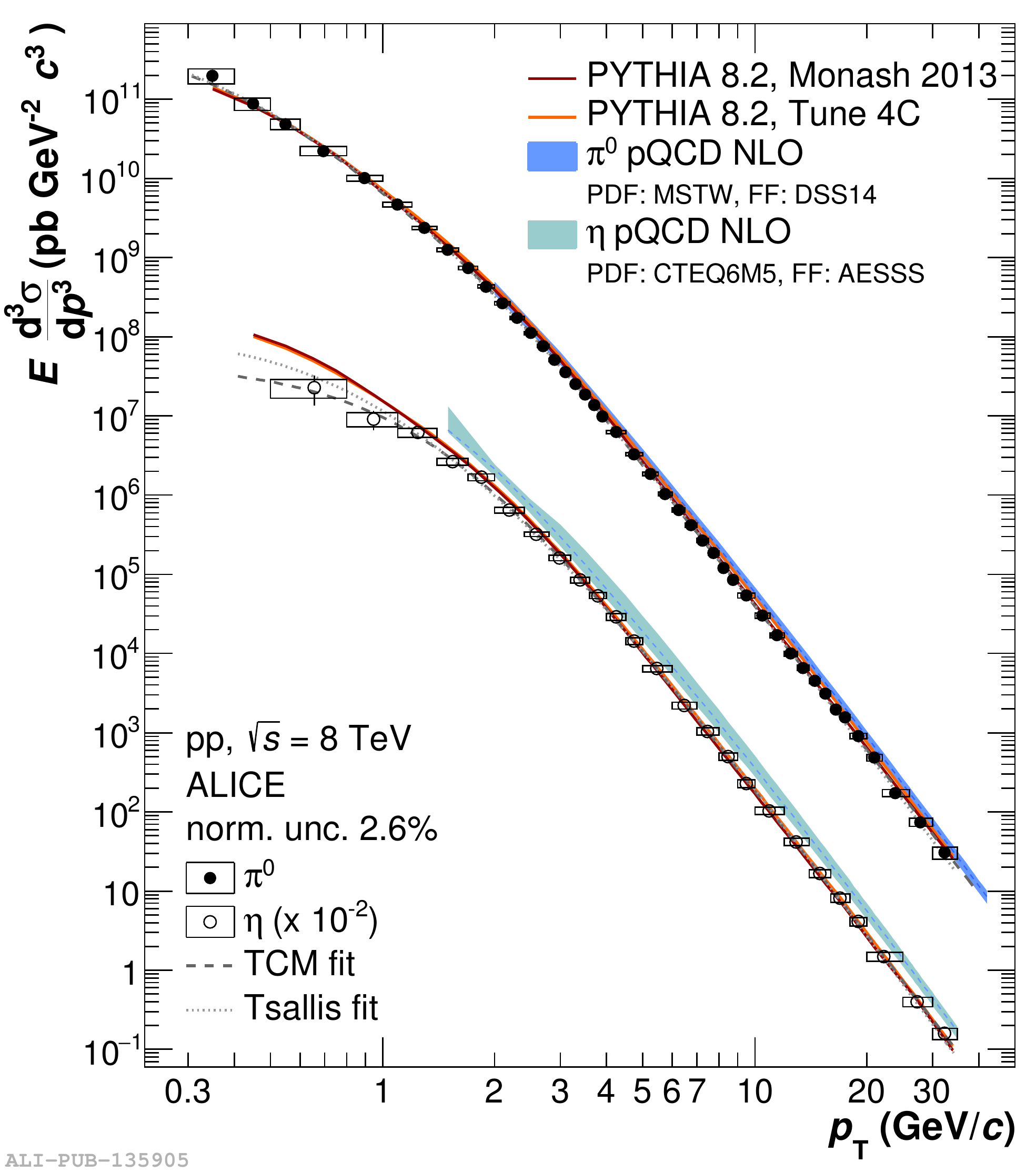}
\includegraphics[width=0.321\textwidth]{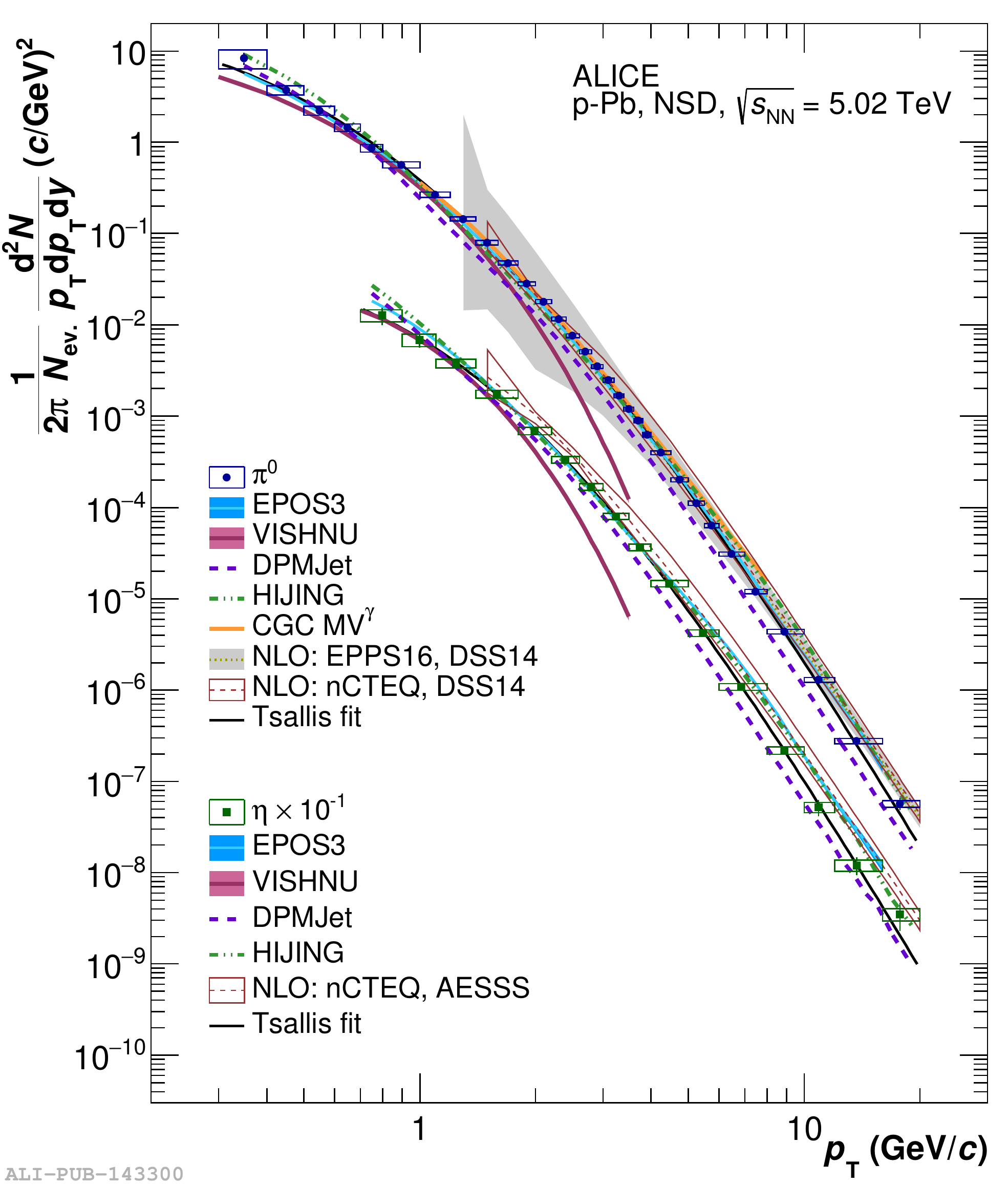}
\includegraphics[width=0.328\textwidth]{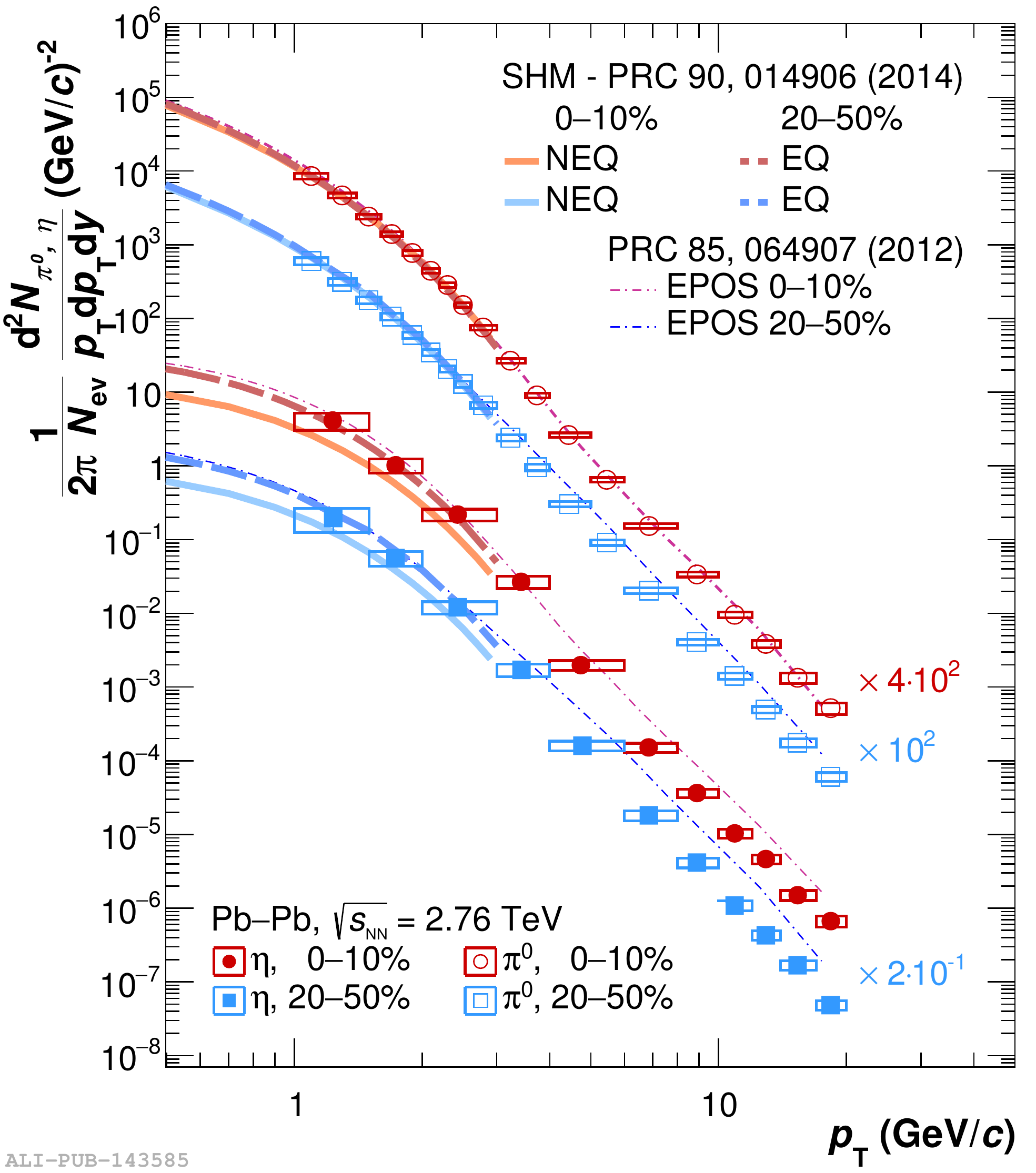}
\vspace*{-0.7cm}
\caption{Neutral meson $p_{\mbox{\tiny T}}$-differential spectra measured in pp collisions at 
$\sqrt{s}$~=~8~TeV (left), in NSD p--Pb collisions at $\sqrt{s_{\mbox{\tiny NN}}}$~=~5.02~TeV 
(middle) and Pb--Pb collisions at $\sqrt{s_{\mbox{\tiny NN}}}$~=~2.76~TeV (right). The ALICE 
data are compared to several theory predictions, detailed and referenced in the respective papers~$^{6,7,8}$.}
\label{fig:NeutralMesonSpectra}
\end{figure}

\begin{figure}[!h]
\centering
\includegraphics[width=0.48\textwidth]{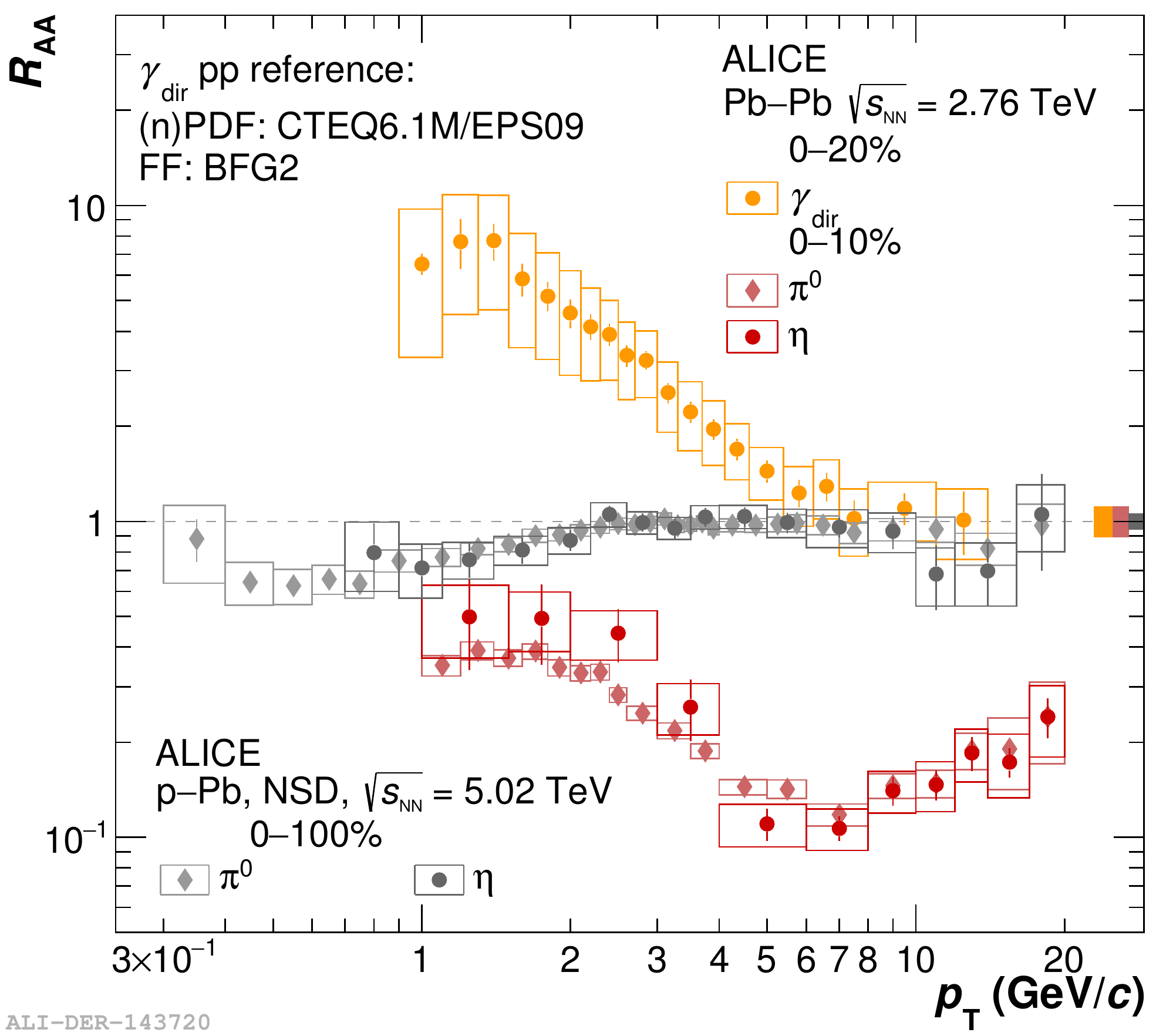}
\vspace*{-0.3cm}
\caption{Nuclear modification factor of direct photons in 0--20\% Pb--Pb collisions at 
$\sqrt{s_{\mbox{\tiny NN}}}$~=~2.76~TeV~$^{9}$ and of neutral mesons in 0--10\% Pb--Pb collisions at 
$\sqrt{s_{\mbox{\tiny NN}}}$~=~2.76~TeV~$^{8}$ and NSD p--Pb collisions at 
$\sqrt{s_{\mbox{\tiny NN}}}$~=~5.02~TeV~$^{7}$.}
\label{fig:RAAphotonmesons}
\end{figure}

Photons are measured in ALICE with the electromagnetic calorimeters (EMCal, PHOS) or 
reconstructing electron-positron pairs coming from photon conversions in the detector 
material~(PCM)~\cite{Abelev:2014ffa}. 
Neutral meson and direct photon measurements are closely connected: the main background for the 
direct photon measurement are the photons from hadron decays, the largest contributions are given by the 
neutral pion and $\eta$ meson decays in two photons. Therefore, for a good direct photon measurement, 
a very precise knowledge of the decay photon background and therefore an accurate measurement of the neutral 
meson contribution is necessary. \\
Neutral pions and $\eta$ mesons have been measured in pp collisions at $\sqrt{s}$~=~0.9, 2.76, 7 
and 8~TeV, in non-single diffractive (NSD) p--Pb collisions at $\sqrt{s_{\mbox{\tiny NN}}}$~=~5.02~TeV 
and in Pb--Pb collisions at $\sqrt{s_{\mbox{\tiny NN}}}$~=~2.76~TeV. Fig.~\ref{fig:NeutralMesonSpectra} 
shows only the most recent results for pp collisions at $\sqrt{s}$~=~8~TeV (left)~\cite{Acharya:2017tlv}, p--Pb 
(middle)~\cite{Acharya:2018hzf} and Pb--Pb collisions (right)~\cite{Acharya:2018yhg}.
\\
These results illustrate ALICE capabilities of measuring identified particles over a large 
transverse momentum range, combining the different reconstruction methods. In 
Fig.~\ref{fig:NeutralMesonSpectra}, the comparison to several theoretical predictions is also 
shown (detailed descriptions and references are given in the respective ALICE papers). While the 
model description of the neutral pion spectra is overall good, the agreement is worse for the 
$\eta$ meson. Thus, the ALICE measurements could help in improving 
the theory if, for example, they are included in the global parameterization of the $\eta$ fragmentation function. \\
Once the neutral mesons have been measured, the spectra are used to estimate 
the total decay photon contribution. This contribution will be then subtracted from the 
inclusive photons to obtain the direct photon signal. The direct photon spectra are composed 
of two types of photons: photon from hard scattering (originating from the same environment as jets), 
dominant at high-$p_{\mbox{\tiny T}}$ and described by NLO pQCD, and photon from thermal emission 
of the QGP and hadronic phase, dominant at low-$p_{\mbox{\tiny T}}$. This second component is used to extract the 
medium characteristics, \emph{e.g.} the emission temperature~\cite{Adam:2015lda}. \\
Fig.~\ref{fig:RAAphotonmesons} shows the neutral meson (in p--Pb and Pb--Pb) and direct 
photon (in Pb--Pb) nuclear modification factors, $R_{\mbox{\tiny AA}}$, as a function of the 
transverse momentum. The $R_{\mbox{\tiny AA}}$ is calculated as the ratio of the yields measured in Pb--Pb 
(p--Pb) collisions over the yields measured in pp collisions, scaled by the number of binary 
collisions. Since no precise pp reference exists for the direct photons, a theoretical calculation is 
used instead, as illustrated in the figure. 
Focusing on the low-$p_{\mbox{\tiny T}}$ direct photon measurement, a large excess above unity can 
be observed. This is the thermal photon component, characteristic of heavy-ion collisions, where the QGP 
forms, which is absent in minimum bias pp collisions. Considering the high-$p_{\mbox{\tiny T}}$ region instead, 
we observe the expected behaviour as mentioned in the introduction. Direct photons are expected to be 
unaffected by the QGP, and in fact the $R_{\mbox{\tiny AA}}$ agrees with unity within uncertainties. On 
the other hand, the hadron yields show a strong suppression, reflecting the energy loss of the parton 
from which they originate in the strongly interacting medium. Moreover, we have the confirmation from 
the p--Pb measurement that the suppression is indeed a final state effect: no such suppression is observed 
for the neutral mesons measured in p--Pb collisions.

\section{Summary}
Jets and photons are complementary probes that can be used to study the characteristics of the QGP in 
Pb--Pb collisions and to judge whether or not this medium is formed in small systems as produced in p--Pb collisions. 
Concerning the latter, no evidence of medium induced effects has been found neither in the jet nor in 
the high-$p_{\mbox{\tiny T}}$ hadron measurements, which are therefore used to set a limit for jet 
quenching in small systems and disentangle initial and final state effects. 
Conversely, the presence of QGP is suggested by the strong suppression of jet and hadron yields 
at intermediate and high-$p_{\mbox{\tiny T}}$ in Pb--Pb collisions and by the excess of direct 
photons at low-$p_{\mbox{\tiny T}}$, consistent with the emission from a hot medium.

\section*{Acknowledgments}
I would like to thank the Heidelberg Graduate School for Fundamental Physics (HGSFP)
for providing the funding that allowed me to give my contribution to the conference.


\section*{References}
\begin{thebibliography}{99}
\bibitem{Abelev:2014ffa} (ALICE Collaboration) B.~Abelev {\it et al.}, \Journal{\IJMP}{29}{1430044}{2014}.
\bibitem{Adam:2015ewa} (ALICE Collaboration) J.~Adam {\it et al.}, \Journal{\PLB}{746}{1}{2015}.
\bibitem{Acharya:2017goa} (ALICE Collaboration) S.~Acharya {\it et al.}, \Journal{\PLB}{776}{249}{2018}.
\bibitem{Adam:2015doa} (ALICE Collaboration) J.~Adam {\it et al.}, \Journal{\JHEP}{09}{170}{2015}.
\bibitem{Acharya:2017okq} (ALICE Collaboration) S.~Acharya {\it et al.}, arXiv/nucl-ex:1712.05603.
\bibitem{Acharya:2017tlv} (ALICE Collaboration) S.~Acharya {\it et al.}, \Journal{\EPJ}{78}{263}{2018}.
\bibitem{Acharya:2018hzf} (ALICE Collaboration) S.~Acharya {\it et al.}, arXiv/nucl-ex:1801.07051.
\bibitem{Acharya:2018yhg} (ALICE Collaboration) S.~Acharya {\it et al.}, arXiv/nucl-ex:1803.05490.
\bibitem{Adam:2015lda} (ALICE Collaboration) J.~Adam {\it et al.}, \Journal{\PLB}{754}{235}{2016}.
\end{thebibliography}
\end{document}